**Conference paper:**
Analytical Investigation by Using the Two-fluid-model to Study the Interfacial Behavior of Air-water Horizontal Stratified Flow

**Authors:**
Hadiyan Yusuf Kuntoro; Deendarlianto; Indarto

**Publication date:**
23 May 2014

**Conference:**
Seminar Nasional Perkembangan Riset dan Teknologi di Bidang Industri ke-20

**Venue:**
Universitas Gadjah Mada, Yogyakarta, Indonesia

**Conference date:**
23 May 2014

# Analytical Investigation by Using the Two-Fluid-Model to Study the Interfacial Behavior of Air-Water Horizontal Stratified Flow

**Hadiyan Yusuf Kuntoro[1], Deendarlianto[2], Indarto[2]**
[1]Fast Track Postgraduate Program of Mechanical Engineering,
Dept. of Mechanical and Industrial Engineering, Faculty of Engineering, Gadjah Mada University.
Jalan Grafika 2 Yogyakarta 55281, Indonesia.
E-mail: hadiyan.y.kuntoro@mail.ugm.ac.id; hadiyan.y.kuntoro@gmail.com

[2]Department of Mechanical and Industrial Engineering, Faculty of Engineering, Gadjah Mada University.
Jalan Grafika 2 Yogyakarta 55281, Indonesia.

**Abstract**

*In the chemical, petroleum and nuclear industries, pipelines are often used to transport fluids from one process site to another one. The understanding of the fluids behavior inside the pipelines is the most important consideration for the engineers and scientists. From the previous studies, there are several two-phase flow patterns in horizontal pipe. One of them is stratified flow pattern, which is characterized by the liquid flowing along the bottom of the pipe and the gas moving above it cocurrently. Another flow patterns are slug and plug flow patterns. This kind of flow triggers the damage in pipelines, such as corrosion, abrasion, and blasting pipe. Therefore, slug and plug flow patterns are undesirable in pipelines, and the flow is maintained at the stratified flow condition for safety reason. In this paper, the analytical-based study on the experiment of the stratified flow pattern in a 26 mm i.d. horizontal pipe is presented. The experiment is performed to develop a high quality database of the stratified two-phase flow pattern. The experimental data were obtained from the visualization data by using a high speed-video camera and were processed by using digital image processing technique. Analytical method of two-fluid-model was used to study the interfacial behavior of the flow. The aim of this study is to validate the previous correlations which are proposed by other researchers. The results show that there are still many significant differences among each other. Hence, better correlation should be proposed in the future. The discussion on the basis of the comparison between the previous correlations with the present analytical data is presented.*

*Keywords: Two-fluid-model, analytical studies, interfacial behavior, stratified flow, two-phase flow.*

## 1. Introduction

In the chemical, petroleum and nuclear industries, pipelines are often used to transport fluids from one process site to another one. The installation of pipelines system is relatively costly, so that the understandings to avoid the damage in pipelines and improve the effectiveness of the fluids transport are important. One of the considerations of the pipelines system design is the fluids that flowing inside it includes the interfacial behavior of the flow. Thus, study regarding the interfacial behavior of the flow is needed.

From the previous studies [Mandhane et al (1974), Weisman et al (1979), Spedding and Nguyen (1980)], there are several two-phase flow patterns in horizontal pipe, for examples are slug and plug flow patterns. These kinds of flow patterns can trigger the damage in pipelines, such as corrosion, abrasion, and blasting pipe. Another flow pattern is stratified flow pattern (Fig. 1), which is characterized by the liquid flowing along the bottom of the pipe and the gas moving above it cocurrently. Ilman and Kusmono (2014) analyzed the internal corrosion in subsea oil pipeline. They concluded that to reduce the incidence of corrosion, the flow should be maintained in the stratified flow pattern.

Some researchers had performed investigations regarding the interfacial behavior of the stratified two-phase flow pattern and proposed their correlations, among others Kowalski (1987), Paras et al (1994), Vlachos et al (1997), and Sidi-Ali and Gatignol (2010). In this paper, several interfacial correlations from them are validated with the present analytical data. The present analytical data are calculated by using the two-fluid-model of Taitel and Dukler (1976). The discussion on the basis of the comparison between the other correlations with the present analytical data is presented.

## 2. Analytical Model

The important interfacial behavior of the stratified two-phase flow pattern is the interfacial shear stress of it. To quantify the interfacial shear stress, first, the interfacial friction factor should be determined. Many researchers [Taitel and Dukler (1976), Kowalski (1987), Paras et al (1994), Vlachos et al (1997), Sidi-Ali and Gatignol (2010)] had proposed their own correlation of it but the results are different among each other.

One of the popular calculations of the interfacial shear stress is by the two-fluid-model considering the momentum balance of each phase, reintroduced by Taitel and Dukler (1976). The model is explained as follows:

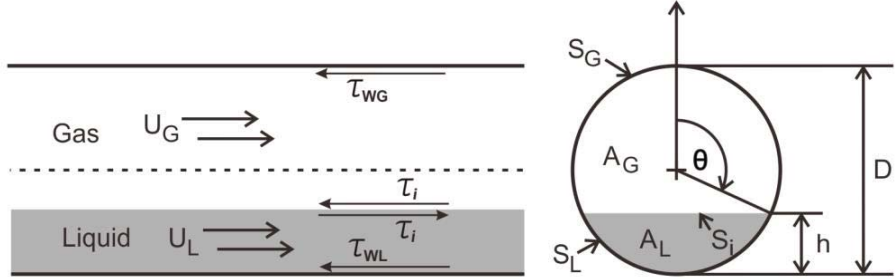

**Fig. 1.** Horizontal stratified two-phase flow model and its geometry parameter.

For steady-state fully-developed gas-liquid flow in horizontal pipe with constant cross-sectional area, momentum balance of each phase:

$$-A_L \left(\frac{dp}{dx}\right) - \tau_{WL} S_L + \tau_i S_i = 0 \qquad (1)$$

$$-A_G \left(\frac{dp}{dx}\right) - \tau_{WG} S_G - \tau_i S_i = 0 \qquad (2)$$

By equating pressure drop in the two phases and assuming that at transition conditions the hydraulic gradient in the liquid is negligible, the results become:

$$\tau_{WG} \frac{S_G}{A_G} - \tau_{WL} \frac{S_L}{A_L} + \tau_i S_i \left(\frac{1}{A_L} + \frac{1}{A_G}\right) = 0 \qquad (3)$$

Then, the shear stresses are evaluated in a conventional manner:

$$\tau_{WL} = f_L \frac{\rho_L U_L^2}{2} \qquad (4)$$

$$\tau_{WG} = f_G \frac{\rho_G U_G^2}{2} \qquad (5)$$

$$\tau_i = f_i \frac{\rho_G (U_G - U_i)^2}{2} \qquad (6)$$

with the liquid, gas, and interfacial friction factors evaluated from

$$f_L = C_L \left(\frac{D_L U_L}{v_L}\right)^{-n} \qquad (7)$$

$$f_G = f_i = C_G \left(\frac{D_G U_G}{v_G}\right)^{-m} \qquad (8)$$

where $D_L$ and $D_G$ are the hydraulic diameter evaluated in the manner as follows:

$$D_L = \frac{4A_L}{S_L} \qquad (9)$$

$$D_G = \frac{4A_G}{S_G + S_i} \qquad (10)$$

The geometry parameter of $S_G$, $S_L$ and $S_i$ are calculated as explained by Kolev (2005):

$$S_G = \theta\, D \qquad (11)$$
$$S_L = (\pi - \theta)\, D \qquad (12)$$
$$S_i = D \sin \theta \qquad (13)$$

The value of $\theta$ is evaluated from Eq. 14. The value of h (film thickness), $A_L$ (liquid holdup/ η) and $A_G$ (void fraction/ α) are obtained from the experimental data.

$$\theta = \cos^{-1}\left(\frac{2h}{D} - 1\right); \quad \theta \; in \; radian \tag{14}$$

Based on the Taitel and Dukler's (1976) work, the coefficients ($C_G$, $C_L$, n, m) were utilized:

$$C_G = C_L = 0.046; \quad n = m = 0.2 \quad \text{(For turbulent flow)} \tag{15}$$
$$C_G = C_L = 16; \quad n = m = 1.0 \quad \text{(For laminar flow)} \tag{16}$$

To identify the flow of each phase (gas & liquid) whether laminar or turbulent, Reynolds number is used to calculate as follows:

$$Re_L = \frac{D_L U_L}{v_L} \tag{17}$$

$$Re_G = \frac{D_G U_G}{v_G} \tag{18}$$

where

$$U_L = \frac{J_L}{\eta} \tag{19}$$

$$U_G = \frac{J_G}{\alpha} \tag{20}$$

In the present investigation, Taitel and Dukler's (1976) analytical prediction was used to find-out the interfacial shear stresses of Kuntoro et al's (2013, 2014) experiment, and the results were compared to the other correlations proposed by Kowalski (1987), Paras et al (1994), Vlachos et al (1997), and Sidi-Ali and Gatignol (2010).

Kowalski (1987) performed an experiment to determine the interfacial shear stress of the stratified two-phase flow in a 50.8 mm i.d. horizontal pipe. For the stratified wavy, the interfacial friction factors were stated in Eq. 21 and Eq. 22 for the stratified smooth.

$$f_i = 7.5 \times 10^{-5}(\eta)^{-0.25} Re_{AG}^{-0.3} Re_{AL}^{0.83} \tag{21}$$
$$f_i = 0.96(Re_{SG})^{-0.52} \tag{22}$$

where

$$Re_{AG} = \frac{D \; U_G}{v_G} \tag{23}$$
$$Re_{AL} = \frac{D \; U_L}{v_L} \tag{24}$$
$$Re_{SG} = \frac{D \; J_G}{v_G} \tag{25}$$
$$Re_{SL} = \frac{D \; J_L}{v_L} \tag{26}$$

Paras et al (1994) and Vlachos et al (1997) also conducted an experiment to determine it, and it was stated in Eq. 27 for Paras et al (1994) and Eq. 28 for Vlachos et al (1997). Paras et al (1994) investigated in a 50.8 mm i.d. horizontal pipe, while Vlachos et al (1997) investigated in a 24 mm i.d. horizontal pipe.

$$f_i = 0.022 + 0.37 \times 10^{-6} Re_{LF} \tag{27}$$
$$f_i = 0.024 \eta^{0.35} Re_{SL}^{0.18} \tag{28}$$

where

$$Re_{LF} = \frac{J_L \; h}{\eta \; v_L} \tag{29}$$

Recently, Sidi-Ali and Gatignol (2010) determined the interfacial shear stress using CFD method and proposed the interfacial friction factor ($f_i$) in Eq. 30.

$$f_i = 0.94 Re_G^{-0.427} \tag{30}$$

**3.   Experimental Data**

To evaluate the equations, the film thickness data are required. The data were obtained from the experiment of Kuntoro et al (2013, 2014) in the range as shown in Table 1 and Fig. 2. The experiment was conducted in a 26 mm i.d. horizontal pipe on the focus of the stratified two-phase flow pattern. Digital image processing technique was used to obtain the liquid film thickness (h) data as well as the liquid holdup (η) data. Detail explanation regarding the experiment and the technique to obtain the film thickness (h) data had reported by Kuntoro et al (2013, 2014).

**Table 1.** Matrix data.

|  | $J_L =$ | | | | | |
|---|---|---|---|---|---|---|
|  | 0.016 m/s | 0.031 m/s | 0.047 m/s | 0.063 m/s | 0.077 m/s | 0.092 m/s |
| $J_G$ = 1.02 m/s | 1 | 2 | 3 | 4 | 5 | 6 |
| $J_G$ = 1.88 m/s | 7 | 8 | 9 | 10 | 11 | 12 |
| $J_G$ = 2.83 m/s | 13 | 14 | 15 | 16 | 17 | 18 |
| $J_G$ = 3.77 m/s | 19 | 20 | 21 | 22 | 23 | 24 |

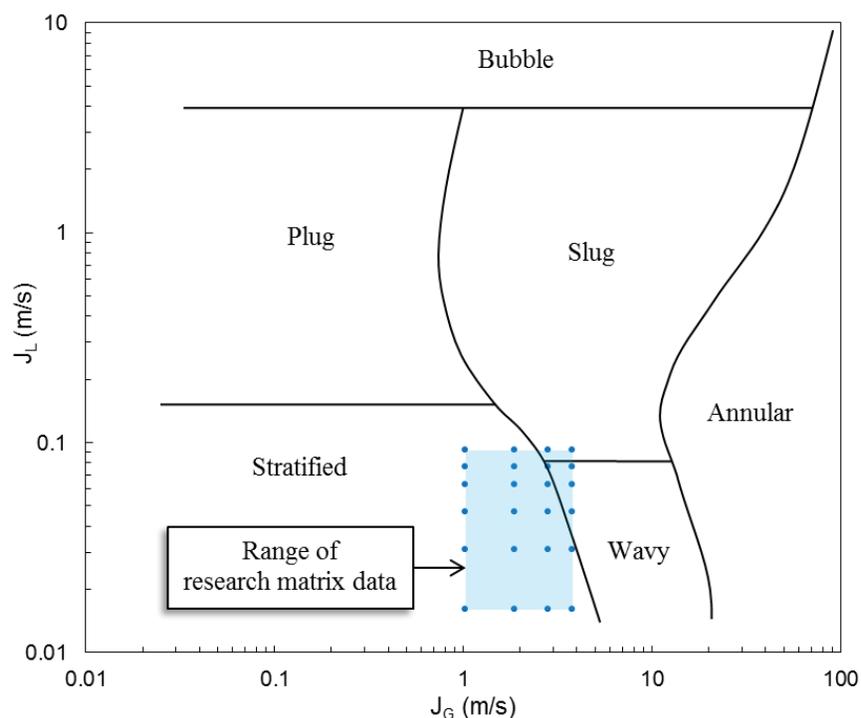

**Fig. 2.** Range of matrix data in Mandhane et al (1974) flow pattern map.

### 4. Results and Discussion

Fig. 3 shows the plots of the interfacial friction factors of the stratified smooth flow pattern at $J_L$ = 0.016 m/s, and stratified wavy flow pattern at $J_L$ = 0.031 m/s. The plots are aimed to investigate the effect of $J_G$ on the interfacial friction factors ($f_i$). Along an increase in $J_G$, the value of $f_i$ does not change significantly.

For the case of the stratified smooth (Fig. 3.a.), the almost similar results of $f_i$ occur at $J_G$ = 1.02 m/s between Kowalski (1987) and Paras et al (1994). The slightly similar results are also shown at $J_G \geq 2.83$ m/s between Paras et al (1994) and Sidi-Ali and Gatignol (2010), and between Kowalski (1987) and the present data.

In the stratified wavy case (Fig. 3.b.), the nearly similar results are shown between Paras et al (1994) and Sidi-Ali and Gatignol (2010), and between Kowalski (1987) and the present data. Both on the stratified smooth and stratified wavy cases, the correlation of Vlachos et al (1997) in Fig. 3 show the distinctive results.

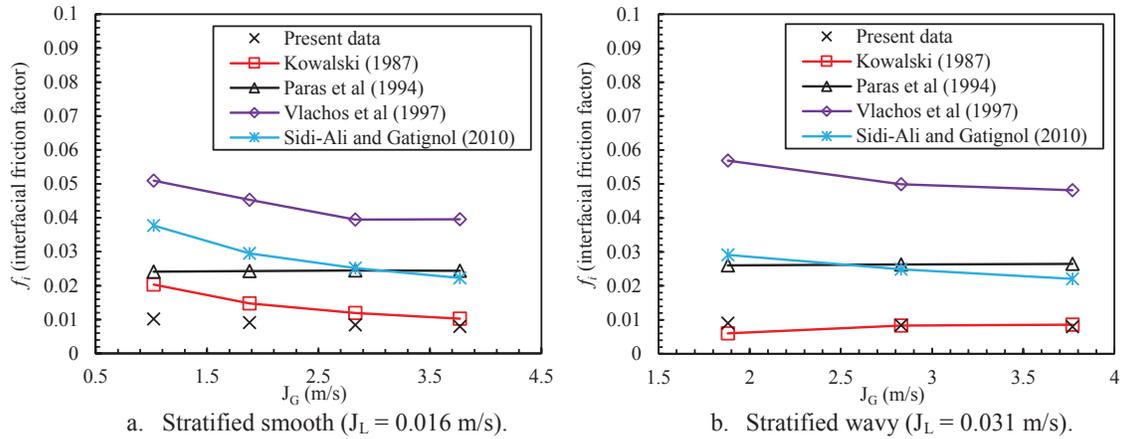

a. Stratified smooth ($J_L$ = 0.016 m/s).     b. Stratified wavy ($J_L$ = 0.031 m/s).
**Fig. 3.** The effect of $J_G$ on the interfacial friction factor in the stratified flow pattern.

Fig. 4 shows the effect of $J_G$ on the interfacial shear stresses. It can be seen that the higher the $J_G$ the higher the interfacial shear stresses ($\tau_i$). Each researcher shows the different correlation results, especially for Vlachos et al (1997). In comparison with other correlations, the present data are the lowest. This is because the Taitel and Dukler's (1976) two-fluid-model only uses analytical prediction method. It does not include the experimental data. Meanwhile, Kowalski (1987), Paras et al (1994), and Vlachos et al (1997) proposed the correlations which were based on their experiment. Different from the others, Sidi-Ali and Gatignol (2010) developed the correlation of $f_i$ ad $\tau_i$ by using the CFD-FLUENT method. Their result shows slightly similar to Paras et al (1994).

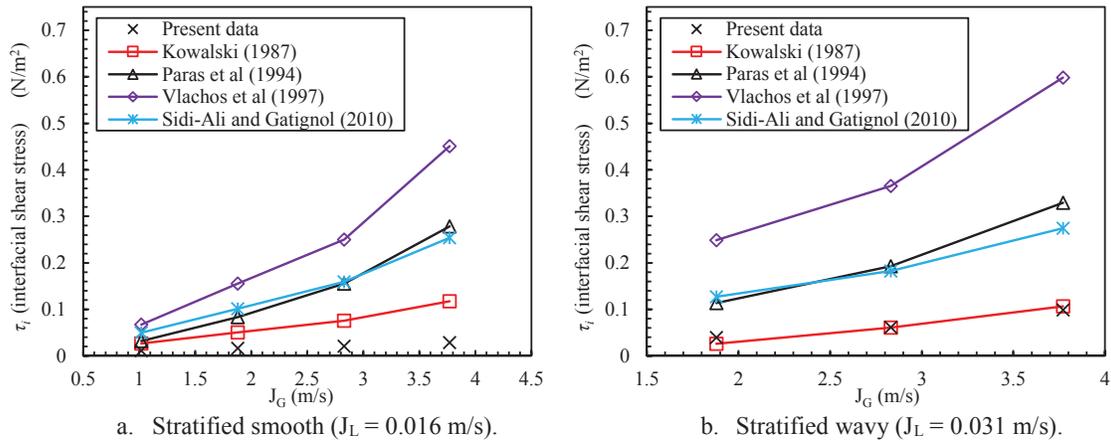

a. Stratified smooth ($J_L$ = 0.016 m/s).     b. Stratified wavy ($J_L$ = 0.031 m/s).
**Fig. 4.** The effect of $J_G$ on the interfacial shear stress in the stratified flow pattern.

In this study, the plot graphs only show the effect of $J_G$ on the interfacial friction factors and interfacial shear stresses due to the lack of the present experimental data. Therefore, further study is needed to investigate the effect of $J_L$ on the interfacial friction factors and the interfacial shear stresses.

## 5. Conclusions

The analytical investigation by using the two-fluid-model to study the interfacial behavior of the stratified two-phase flow pattern, which is compared to the interfacial correlations from Kowalski (1987), Paras et al (1994), Vlachos et al (1997), Sidi-Ali and Gatignol (2010), is presented. In the present work, the film thickness data are obtained from the experimental data of Kuntoro et al (2013, 2014). The analytical prediction method of two-fluid-model reintroduced by Taitel and Dukler (1976) is used to calculate the interfacial friction factors ($f_i$) and the interfacial shear stresses ($\tau_i$). In this investigation, the calculation of $f_i$ and $\tau_i$ show the lowest results rather than the results by Kowalski (1987), Paras et al (1994), Vlachos et al (1997), and Sidi-Ali and Gatignol (2010). It can be concluded that the higher the $J_G$ the higher the interfacial shear stresses of the flow. Many dissimilar results can be important background to conduct the future works regarding the interfacial friction factors and the interfacial shear stresses.

**Acknowledgments**

The authors gratefully acknowledge the financial support of 'Hibah Penelitian Unggulan Perguruan Tinggi (contract number LPPM-UGM/1448/LIT/2013)' from Indonesia Directorate General for Higher Education (DIKTI), Ministry of Education and Culture, Republic of Indonesia. The first author acknowledges the support of Fast Track Scholarship Program from Indonesia Directorate General for Higher Education (DIKTI), Ministry of Education and Culture, Republic of Indonesia.

**Nomenclature**

| | | | | | |
|---|---|---|---|---|---|
| $\frac{dp}{dx}$ | pressure drop (Pa/m) | S | perimeter (m) | *Subscripts* | |
| A | cross-sectional area (m$^2$) | U | actual velocity (m/s) | A | actual |
| C,m,n | constant | α | void fraction | G | gas |
| D | diameter (m) | η | liquid holdup | I | interfacial |
| f | friction factor | θ | angle (radian) | L | liquid |
| h | film thickness (m) | ρ | density (kg/m$^3$) | LF | liquid fraction |
| J | superficial velocity (m/s) | τ | shear stress (Pa) | S | superficial |
| Re | Reynolds number | υ | kinematic viscosity (m$^2$/s) | | |